\documentclass[tightenlines,showpacs,prd,floatfix,preprintnumbers,amsmath,amssymb,nofootinbib,superscriptaddress]{revtex4-1}
\DeclareMathAlphabet{\mathpzc}{OT1}{pzc}{m}{it}
\usepackage{graphicx}
\usepackage{amsfonts}
\usepackage[normalem]{ulem}
\usepackage{color}
\usepackage{bm}
\usepackage{comment}
\usepackage{ulem}

\def\up#1{\raise1mm\hbox{$\!\!^{#1}$}} 
\def\upp#1{\raise2mm\hbox{$\!\!\!\!^{#1}$}} 
\newcommand{\Deqn}[1]{{Eq.~(\ref{#1})}}

\newcommand{\beq}{\begin{equation}}
\newcommand{\be}{\begin{equation}}
\newcommand{\eeq}{\end{equation}}
\newcommand{\ee}{\end{equation}}
\newcommand{\bea}{\begin{eqnarray}}
\newcommand{\eea}{\end{eqnarray}}
\newcommand{\bal}{\begin{align}}
\newcommand{\eal}{\end{align}}

\newcommand{\beaa}{\begin{eqnarray*}} 
\newcommand{\eeaa}{\end{eqnarray*}}

\newcommand{\bsube}{\begin{subequations}}
\newcommand{\esube}{\end{subequations}}

\begin{document}

\title{Finding high-order {\em analytic} post-Newtonian parameters from a high-precision {\em numerical} self-force calculation}

\author{Abhay G. Shah} 
\email{a.g.shah@soton.ac.uk}
\affiliation{Dept of Particle Physics $\&$ Astrophysics, Weizmann Institute of Science, Rehovot 76100, Israel}
\affiliation{School of Mathematics, University of Southampton, Southampton SO17 1BJ, United Kingdom}

\author{John L. Friedman} 
\email{friedman@uwm.edu}
\affiliation{Center for Gravitation and Cosmology, Department of Physics, University of Wisconsin--Milwaukee, P.O. Box 413, Milwaukee, Wisconsin 53201, USA}

\author{Bernard F. Whiting} 
\email{bernard@phys.ufl.edu}
\affiliation{Department of Physics, P.O. Box 118440, University of Florida, Gainesville, Florida 32611-8440}
\affiliation{$\mathcal{G}\mathbb{R}\varepsilon{\mathbb{C}}\mathcal{O}$,
   Institut d'Astrophysique de Paris --- UMR 7095 du CNRS,
   \\ Universit\'e Pierre \& Marie Curie, 98\textsuperscript{bis}
   boulevard Arago, 75014 Paris, France}


\begin{abstract}

We present a novel  analytic extraction of high-order post-Newtonian (pN) parameters
that govern quasi-circular binary systems.  Coefficients in the pN expansion 
of the energy of a binary system can be found from corresponding coefficients 
in an extreme-mass-ratio inspiral (EMRI) computation of the change 
$\Delta U$ in the redshift factor of a circular orbit at fixed angular velocity.  
Remarkably, by computing this essentially gauge-invariant quantity to accuracy 
greater than one part in $10^{225}$, and by assuming that a subset of pN 
coefficients are rational numbers or products of $\pi$ and a rational, 
we obtain the exact analytic coefficients.   We find 
the previously unexpected result that the post-Newtonian expansion of 
$\Delta U$ (and of the change $\Delta\Omega$ in the angular velocity at 
fixed redshift factor) have conservative terms at half-integral pN order 
beginning with a 5.5 pN term. This implies the existence of a corresponding 
5.5 pN term in the expansion of the energy of a binary system.  

Coefficients in the pN series that do not belong to the subset just described
are obtained to accuracy better than 1 part in $10^{265-23n}$ at $n$th pN order.  
We work in a radiation gauge, finding the radiative part of the metric 
perturbation from the gauge-invariant Weyl scalar $\psi_0$ via a Hertz potential.  We use mode-sum renormalization, and find high-order renormalization coefficients by matching 
a series in $L=\ell+1/2$ to the large-$L$ behavior of the expression for $\Delta U$.
The non-radiative parts of the perturbed metric associated with changes 
in mass and angular momentum are calculated in the Schwarzschild gauge. 

\end{abstract}

\maketitle

\section{Introduction}

     The principal approximation methods used to compute the inspiral of 
compact binary systems are the post-Newtonian expansion, in which an 
orbital angular velocity $M\Omega$ serves as the expansion parameter; and the self-force 
or extreme-mass-ratio-inspiral (EMRI) approach, in which the small parameter 
is the mass ratio ${\mathfrak m}/M$ of the binary's two components.  Previous work 
by Blanchet et al. \cite{Blanchetetal,BDTW11} used an overlapping regime 
where both approximations are valid to check the consistency of the  
renormalization methods used in the two approaches and to find numerical 
values of pN coefficients at orders beyond the reach of current analytical 
work.  

In the present paper, by working with much higher numerical accuracy -- 
maintaining precision of at least one part in $10^{225}$ in an EMRI computation 
of the perturbed orbital frequency and redshift factor,  
and by considering orbits at much larger separation -- with orbital radii 
extending to $10^{30}M$, we obtain two surprising results not seen in the 
previous study:  
\begin{itemize}
\item (1)  A subset of the pN parameters in lower-order 
analytical work had been found to be either rationals $m/n$ or 
to be sums of rationals multiplied by powers of $\pi$, the Euler 
constant $\gamma$ and square roots of integers. Our high precision allows us to 
extract the exact {\em analytical} form of the subset of coefficients 
that are rationals or products of the form rational$\times \pi$ from 
our {\em numerical} values up to 10 pN order, corresponding to corrections 
smaller by $(v/c)^{22}$ that the Newtonian value.  
\item (2)  In a pN expansion,
conservative terms (terms even under the interchange of outgoing and
ingoing radiation) are initially encountered at integral pN orders;  
dissipative terms (odd under the interchange of outgoing and
ingoing) first enter at 2.5 pN order. 
At higher order, dissipative terms can occur at
either integral or half-integral order, depending on the details \cite{BlanchetDamour88}, 
while conservative terms enter at each integral order.  We find 
that conservative terms of 5.5 pN order appear in the expression 
for the redshift at fixed angular velocity (and thus in the 
expressions for the angular velocity at fixed redshift and in the 
expression for the energy of an orbit with given angular velocity). 
These quantities are conservative, and the presence of 5.5 pN terms
was unexpected.
\end{itemize}

  The work reported here involves a binary system that, at zeroth order 
in the mass ratio, is described by a test particle in a circular geodesic
about a Schwarzschild black hole.  At first order in ${\mathfrak m}/M$, the orbital 
parameters are altered by a metric perturbation $h_{\alpha\beta}$ produced
by the orbiting particle: The perturbed motion can be described by saying 
that the particle moves on a circular geodesic of the metric 
$g_{\alpha\beta}+h^{\rm ren}_{\alpha\beta}$, where $h^{\rm ren}_{\alpha\beta}$ is the 
renormalized metric perturbation.  The perturbed spacetime is helically 
symmetric, with a helical Killing vector $k^\alpha$ that is tangent 
to the particle's 4-velocity,
\be
     u^\alpha = U k^\alpha.
\ee
The constant of proportionality $U$ is termed the redshift factor (first introduced by Detweiler \cite{detweiler08}),
and can be thought of as a contribution to the redshift, measured from 
the perturbed orbit of the mass $m$, that is independent of the 
internal geometry of the mass.  With the perturbed spacetime 
chosen so that the perturbed and unperturbed helical Killing vectors 
coincide, the change in $\Delta U$ at fixed angular velocity $\Omega$ 
has the form 
\be
\Delta U = -U \frac12 h^{\rm ren}_{\alpha\beta}u^\alpha u^\beta =: -UH^{\rm ren},
\label{DeltaU}\ee      
and it is invariant under gauge transformations generated by helically 
symmetric gauge vectors. 

A pN expansion of $\Delta U$, written in terms of a dimensionless $R:=(M\Omega)^{-2/3}$,  has the form
\be
   \Delta U = -\frac1R + \sum_{n=1}\alpha_n\frac1{R^{n+1}} 
		+ \sum_{n=4} \beta_n\frac{\log R} {R^{n+1}} 
		+ \sum_{n=7} \gamma_n\frac{\log^2 R}{R^{n+1}} 
		+ \sum_{n=10}\delta_n\frac{\log^3 R}{R^{n+1}}
		+\cdots, 
\label{eq:DeltaU}\ee 
where the post-Newtonian order $n$ can take half-integral 
as well as integral values, starting at $\alpha_{5.5}$ and $\beta_{8.5}$.
That is, integral values of coefficients of $\log^k R/R^{n+1}$ start at 
pN order $n=3k+1$; half-integral values appear to start at $n=3k+5.5$, 
but we do not carry our numerical expansion far enough to find the 
first half-integral value for $k=2$ ($\gamma_{11.5}$) or for larger $k$.    
We compute $\Delta U$ at a set of radii extending to $10^{30} M$ and match to 
a series of this form.   As noted in the abstract, the high numerical accuracy 
of $\Delta U(r)$ allows us to find the coefficients $\alpha_n, \beta_n$, 
and $\gamma_n$ with a precision at least as high as one part in $10^{265-23n}$.
At each pN order, we find that the coefficient of the highest 
occurring power of $\log R$ is rational when $n$ is an integer; and it 
has the form rational$\times \pi$ when $n$ is a half-integer.  The 
remaining coefficients for a given value of $n$ are not of this form. 

Because the presence of $\alpha_{5.5}$ and higher-order 
half-integral coefficients was not expected, we performed an elaborate 
set of checks.  Our calculations were carried out in a radiation gauge, 
but, we repeated the entire numerical calculation of $\Delta U$ in a 
Regge-Wheeler gauge, obtaining numerical agreement to {
368} places of accuracy, that is, 
{the retarded values of $h_{uu}$ for each $\ell$-mode in the RG and RWZ gauges agree to more than 368 digits)}.  This serves as a demanding test of both the numerical 
code and of the analytical computation on which it is based.  Because 
the numerical calculation is performed in Mathematica, the comparison 
is also a check of Mathematica's claimed numerical precision.  
Adrian Ottewill and Marc Casals kindly used their codes to perform 
an independent radiation-gauge computation 
to compare 
with ours at double-precision accuracy for small R. 
{Specifically, for the $s=\ell=m=2$ term, we compared our values of the invariant, $
A_{lm}{R_HR_{\infty}}
$ (see \Deqn{eqAlm} below), at $r/M=10^3,10^6$}.  Finally, we analytically 
computed $\alpha_{5.5}$ (see Sect.~\ref{exact:subsect}).    

In Sect.~\ref{sec2} we briefly review the calculation of the renormalized 
$\Delta U$ in a modified radiation gauge.   In Sect.~\ref{sec3} we present 
the results of matching a sequence of values $\Delta U(r)$ to a series of the 
form (\ref{eq:DeltaU}).  

We work in gravitational units ($G=c=1$) and use signature $+---$ to conform 
to Newman-Penrose conventions.     

\section{Review of $\Delta U$ computation}
\label{sec2}

We consider a particle of mass $\mathfrak{m}$ orbiting a   
Schwarzschild black hole of mass $M$.  At zeroth order in $\mathfrak{m}/M$,  
the trajectory is a circular orbit.  In Schwarzschild coordinates, its 
angular velocity is $\Omega = \sqrt{M/r_0^3}$, and its 4-velocity is 
given by
\be
u^\alpha = U (t^\alpha + \Omega \phi^\alpha),\quad \mbox{with}\ \ 
U = u^t = \frac{1}{\sqrt{1-3M/r_0}}.
\ee
We compute the change $\Delta U$ at first-order in $\mathfrak{m}/M$ in 
a modified radiation gauge, as detailed in \cite{sf4}.  We briefly 
review the formalism here, noting first that Eq.~(\ref{eq:DeltaU}) 
for $\Delta U$ involves a single component $H^{\rm ren}$ of the renormalized 
metric perturbation. 

For multipoles with $\ell\geq 2$, the metric perturbation can be 
found in a radiation gauge from the the spin-2 retarded Weyl scalar, 
$\psi_0$, which has the form \cite{sf2, sf3, sf4}, 
\bea
\psi_0(x) &=& \psi_0^{(0)}+\psi_0^{(1)}+\psi_0^{(2)},
\label{eq:psiGT}\eea
with
\bsube\bea
\psi_0^{(0)} &=& 4\pi {\mathfrak m} u^t \frac{\Delta_0^2}{r_0^2}\sum_{\ell m}A_{\ell m}[(\ell-1)\ell(\ell+1)(\ell+2)]^{1/2}R_{\rm H}(r_<)R_\infty(r_>){}_2Y_{\ell m}(\theta,\phi)\bar{Y}_{\ell m}\left(\frac{\pi}{2},\Omega t\right), \\
\psi_0^{(1)} &=& 8\pi i{\mathfrak m} \Omega u^t \Delta_0 \sum_{\ell m} A_{\ell m}[(\ell-1)(\ell+2)]^{1/2}
	{}_2Y_{\ell m}(\theta,\phi){}_1\bar{Y}_{\ell m}\left(\frac{\pi}{2},\Omega t\right) \times  \nonumber\\
& & \quad \Bigl\{[im\Omega r_0^2 + 2 r_0]R_{\rm H}(r_<)R_\infty(r_>) 
	+ \Delta_0[R_{\rm H}'(r_0)R_\infty(r)\theta(r-r_0) 	 + R_{\rm H}(r)R_\infty'(r_0)\theta(r_0-r)]\Bigr\},\\
\psi_0^{(2)} &=& -4\pi {\mathfrak m}\Omega^2 u^t \sum_{\ell m} A_{\ell m}
	{}_2Y_{\ell m}(\theta,\phi){}_2\bar{Y}_{\ell m}\left(\frac{\pi}{2},\Omega t\right) \times 
\nonumber\\ 
& & \biggl\{[30r_0^4 - 80Mr_0^3 + 48M^2r_0^2 - m^2\Omega^2 r_0^6 -2\Delta_0^2 - 24\Delta_0 r_0(r_0-M)+ 6im\Omega r_0^4(r_0-M)]
	R_{\rm H}(r_<)R_\infty(r_>)
 \nonumber\\ 
& & \qquad  + 2(6r_0^5 - 20Mr_0^4 + 16M^2r_0^3 - 3r_0\Delta_0^2 + im\Omega \Delta_0 r_0^4)[R_{\rm H}'(r_0)R_\infty(r)\theta(r-r_0) 
 + R_\infty'(r_0)R_{\rm H}(r)\theta(r_0-r)] re\nonumber\\ 
& & \qquad + r_0^2\Delta_0^2[R_{\rm H}''(r_0)R_\infty(r)\theta(r-r_0) + R_\infty''(r_0)R_{\rm H}(r)\theta(r_0-r)+\textrm{W}[R_{\rm H}(r),R_\infty(r)]\delta(r-r_0)]\Biggr\},
\eea\esube 
where $\Delta = r^2 - 2Mr$; the functions $R_H$ and $R_\infty$ (indices $\ell,m$ are suppressed) are the solutions to the homogenous radial Teukolsky equation that are ingoing and outgoing at the future event horizon and null infinity, respectively, and a prime denotes their derivative with respect to $r$; $\textrm{W}[R_{\rm H}(r),R_\infty(r)] = R_{\rm H} R_\infty^\prime - R_\infty R_{\rm H}^\prime $;  and the quantities $A_{\ell m}$,
given by 
\be
 A_{\ell m} := \frac{1}{\Delta^3 \textrm{W}[R_{\rm H}(r),R_\infty(r)]},
\label{eqAlm}
\ee
are constants, independent of $r$.  
The functions $R_H$ and $R_\infty$ are calculated to more than 350 digits of accuracy using expansions in terms of hypergeometric functions given in \cite{MST}, namely
\begin{align}
R_H &= e^{i\epsilon x}(-x)^{-2-i\epsilon}\sum_{n=-\infty}^{\infty}a_n F(n+\nu+1-i\epsilon,-n-\nu-i\epsilon,-1-2i\epsilon;x), 
\label{RH}\\
R_\infty &= e^{i z}z^{\nu-2} \sum_{n=-\infty}^{\infty} (-2z)^n b_n U(n+\nu+3-i\epsilon,2n+2\nu+2;-2iz).
\label{radialfunctions}
\end{align}
where $x = 1-\frac{r}{2M}$, $\epsilon = 2 M m\Omega$ and $z = -\epsilon x$. We refer the reader to \cite{MST,LivRev} for the derivation of $\nu$ (the renormalized angular momentum), and the coefficients $a_n$ and $b_n$. Here $F$ and $U$ are the hypergeometric and the (Tricomi's) confluent hypergeometric functions. 

The computation of the spin-weighted spherical harmonics $_sY_{\ell,m}(\theta,\phi)$ 
{is done analytically using \cite{sf3}}.   

Once $\psi_0$ is computed, the components of the metric perturbation are found 
from Hertz potential, $\Psi$, whose angular harmonics are related to those of 
$\psi_0$ by an algebraic equation,
\beq
\Psi_{\ell m} = 8 \frac{(-1)^m (\ell+2)(\ell+1)\ell(\ell-1)\bar\psi_{\ell,-m}
	+ 12 i m M \Omega \psi_{\ell m} }{ [(\ell+2)(\ell+1)\ell(\ell-1)]^2 + 144 m^2 M^2 \Omega ^2}
\eeq
where $\Psi = \sum_{\ell,m}\Psi_{\ell m}(r){\,}_2Y_{\ell m}(\theta,\phi)e^{-im\Omega t}$ and $\psi_0 = \sum_{\ell,m}\psi_{\ell m}(r){\,}_2Y_{\ell m}(\theta,\phi)e^{-im\Omega t}$. 
The components of the metric along the Kinnersley tetrad are 
\bea
h_{\bf 11} &=& \frac{r^2}{2}(\bar{\eth}^2\Psi + \eth^2\overline\Psi), \\ 
h_{\bf 33} &=& r^4\left[\frac{\partial_t^2 -2f\partial_t\partial_r+f^2\partial_r^2}{4} - \frac{3(r-M)}{2r^2}\partial_t
+ \frac{f(3r-2M)}{2r^2}\partial_r + \frac{r^2-2M^2}{r^4}\right]\Psi, \\
h_{\bf 13} &=& -\frac{r^3}{2\sqrt{2}}\left(\partial_t-f\partial_r-\frac{2}{r}\right)\bar{\eth}{\Psi},
\eea
where $f = \Delta/r^2$ and the operators $\eth$ and $\bar{\eth}$, acting on a spin-s quantity, $\eta$, 
are given by
\begin{eqnarray}
\eth\eta &=& -\left(\partial_\theta+i\csc\theta\partial_\phi-s\cot\theta\right)\eta, 
\label{green_eth_b}
\nonumber\\
\bar{\eth}\eta&=&-\left(\partial_\theta-i\csc\theta\partial_\phi+s\cot\theta\right)\eta.
\label{green_eth_bar_b}
\end{eqnarray}

The metric recovered from $\psi_0^{\rm ren}$ above only specifies the radiative part of the perturbations ($\ell \ge 2$) and the full metric reconstruction requires one to take into account the change in mass and angular momentum of the Schwarzschild metric and are associated with $\ell=0$ and $\ell=1$ harmonics, respectively. The contribution to the full $H$ from the change in mass ($H_{\delta M}$) and angular momentum ($H_{\delta J}$) of the Schwarzschild metric are given by (see Eqs.~(137, 138) of \cite{sf4})
\begin{align}
H_{\delta M} & = \frac{\mathfrak{m}(r_0-2M)}{r_0^{1/2}(r_0-3M)^{3/2}}, \\
H_{\delta J} &= \frac{-2M\mathfrak{m}}{r_0^{1/2}(r_0-3M)^{3/2}}.
\end{align}

The renormalization of $H$ is described in detail in \cite{sf2,sf3,sf4}. 
The related quantity $\Delta \Omega$ that gives the $O(\mathfrak m)$ change in the 
the angular velocity of a trajectory at fixed redshift factor is 
\be
\Delta\Omega = -\frac1{u_{\phi}u^t}H^{\rm ren} = \frac{\Delta U}{u_\phi u^{t^2}}.
\ee

\section{Results}
\label{sec3}

In this section we present the pN-coefficients of $\Delta U$. Prior to this work, the following analytical coefficients were known \cite{Blanchetetal,BiniDamour}:
\begin{align} \label{knownpN}
\Delta U = &\frac{-1}{R} + \frac{-2}{R^2} + \frac{-5}{R^3} + \frac{-3872+123\pi^2}{96R^4}  + \frac{-592384 - 196608\gamma + 10155\pi^2 - 393216\log(2)}{7680R^5}\nonumber\\
& + \frac{64 \log(R)}{5R^5} + \frac{-956\log(R)}{105R^6}
\end{align}
We calculate $\Delta U$ for a set of $R$-values from $1\times, 3\times, 5\times, 8\times$$10^{20}$ to $10^{29}$ in logarithmic intervals of 10 with an accuracy of one part in $10^{227}$ for $R=10^{20}$, $10^{242}$ for $R=10^{25}$ and $10^{252}$ for $R=10^{30}$. We then match this data to a pN-series to extract the unknown coefficients. In doing so, we find non-zero half-integer ($n.5$) pN coefficients that come from the tail-of-tail terms in pN-computations \cite{BFW5.5}. To confirm its presence we analytically calculated the 5.5pN term -- the coefficient of $1/R^{6.5}$ and found that it agreed with the numerically extracted coefficient to 113 significant digits.
(The analytic calculation is described briefly below.)    
The high accuracy of the numerically extracted coefficients, however, allows us to extract their exact analytical expressions, {\em without an analytic calculation}. For example, the numerically extracted value of the 6-pN $\log$-term is
\begin{align}
-90.&398589065255731922398589065255731922398589065255731922\nonumber\\
&3985890652557319223985890652557319223985890485251879955\cdots 
\end{align}
More than five repetition cycles of the string $398589065255731922$ tells us that it is the 
rational number $-51256/567$. In a similar fashion we extract analytical values of other coefficients making the pN-series of analytically known coefficients the following,
\begin{align}\label{newpN}
\Delta U_{\mbox{analytically known}} = &\frac{-1}{R} + \frac{-2}{R^2} + \frac{-5}{R^3} + \frac{-3872+123\pi^2}{96R^4}  + \frac{-592384 - 196608\gamma + 10155\pi^2 - 393216\log(2)}{7680R^5}\nonumber\\
& + \frac{64 \log(R)}{5R^5} + \frac{-956\log(R)}{105R^6} + \frac{-13696\pi}{525R^{6.5}} + \frac{-51256\log(R)}{567R^7} + \frac{81077\pi}{3675R^{7.5}} + \frac{27392\log^2(R)}{525R^8}\nonumber\\
& + \frac{82561159\pi}{467775R^{8.5}} + \frac{-27016\log^2(R)}{2205R^9} + \frac{-11723776\pi\log(R)}{55125R^{9.5}} + \frac{-4027582708\log^2(R)}{9823275R^{10}} \nonumber\\
&+ \frac{99186502\pi\log(R)}{1157625R^{10.5}} + \frac{23447552\log^3(R)}{165375R^{11}}.
\end{align}
A rational number with fewer than ten digits in its numerator and in its denominator is determined by the first eleven digits in its decimal expansion; thus if 
one assumes that the rationals occurring in the coefficients of
(\ref{newpN}) have this character, they are uniquely determined by 
the numerical accuracy.  Without the assumption, the probability that the 
first $n$ digits in a decimal representation of a randomly chosen number
will match a rational with  
$n_n$ and $n_d$ digits in numerator and denominator is less than $10^{n_n+n_d-n}$, when $n>n_n+n_d$.

   
After using the above analytical coefficients, a numerical fit for the other numerical coefficients in (\ref{eq:DeltaU}) gives the values listed in Table \ref{table}.

\begin{center}
\begin{table}[h!]
  \begin{tabular}{|l|l|}
    \hline
                 \multicolumn{1}{|p{1.5cm}|}
                 {\centering Coefficient}
                & \multicolumn{1}{|p{11.7cm}|}
                 {\centering Numerical value}  \\
    \hline
\hspace{2mm}
$\alpha_5$ & \hspace{2mm}-243.17681446467430758729358896693800234737272817232786539528868308827\\
	   &\hspace{2mm}\phantom{xxxx}94813055787844008820951887564926056965827710452637773038028704808 \footnote{See {\bf Note added} at the end of the
paper.
}\\
\hspace{2mm}
$\alpha_6$ &\hspace{2mm} -1305.0013810787096557410900682717136851595808847394760333078920251334\\
		&\hspace{2mm}\phantom{xxxxx}98776905927112179825227138960576902431854$^\textrm{ a}$\\
\hspace{2mm}
$\alpha_7$ &\hspace{2mm} -6343.8744531990306527270512066053061390446046295187692031581328657892\\
	   &\hspace{2mm}\phantom{xxxxx}063930482892366\\
\hspace{2mm}
$\alpha_8$ &\hspace{2mm} -11903.4729472013044159758685624140826902285745341620173222629\\
\hspace{2mm}
$\alpha_{8.5}$ &\hspace{2mm} -8301.37370829085581136384718573193317705504946743\\
\hspace{2mm}
$\alpha_9$ &\hspace{2mm} -32239.6275950925564123677060345920962 \\
\hspace{2mm}
$\alpha_{9.5}$ &\hspace{2mm} -10864.625586706244075245767\\
\hspace{2mm}
$\alpha_{10}$ &\hspace{2mm} -221316.52514302\\
\hspace{2mm}
$\alpha_{10.5}$ &\hspace{2mm} 6.035$\times10^4$\\
\hspace{2mm}
$\beta_7$ &\hspace{2mm} 536.40521247102428687178953947503891127020626955232120792788336024036\\
&\phantom{xxxx}873632676613183$3^\textrm{ a}$\\
\hspace{2mm}
$\beta_8$ &\hspace{2mm} 1490.5550856958907438011974098988395166992724311135937950474$7^\textrm{ a}$\\
\hspace{2mm}
$\beta_9$ &\hspace{2mm} -3176.929181153969206392338832692666088 \\
\hspace{2mm}
$\beta_{10}$ &\hspace{2mm} -7358.271055677\\
\hspace{2mm}
$\beta_{10.5}$ &\hspace{2mm}  5013.2 \\
\hspace{2mm}
$\gamma_{10} $ &\hspace{2mm} 2105.92718670257 \\
                                                \hline
  \end{tabular}
\caption{Numerical values of the coefficients in the expansion (\ref{DeltaU}) of $\Delta U$ 
for which analytic expressions could not be inferred. }
 \label{table} \end{table}
\end{center}

{\subsection{Exact, Analytic 5.5pN value}
\label{exact:subsect}
As mentioned above, as a check on the work, we analytically compute the 5.5pN term.
To do so, we use the fact that the renormalization parameters that characterize
the singular part of $H^\textrm{ret}$ have \emph{no} {n.5pN} terms: The pN 
expansion of $H^\textrm{sing}$ does not include half-integer powers of $1/R$. 
Studying the pN-expansion of the first few multipoles of $H^\textrm{ret}$, we find   that the 5.5pN term comes only from the $\ell=2,\,m=\pm2$ multipole of $H^\textrm{ret}$. That is, the numerical coefficient of the 5.5-pN term we obtain by matching $H^\textrm{ren}$ coincides exactly with that obtained by matching the sum of 
the $\ell=2,\,m=\pm2$ multipoles of $H^\textrm{ret}$ to a pN-series. 
The analytic calculation was thus restricted to the $\ell=2,\,m=\pm2$ multipoles
of $R_{H}^{(p)}R_{\infty}^{(q)}/(R_{H}R_{\infty}^{\prime}-R_{H}^{\prime}R_{\infty})$ (where $p$ and $q$, the number of radial derivates, each run from 0 to 2). We use 
the hypergeometric series Eqs.~(\ref{RH}) and (\ref{radialfunctions}) 
to express each of these functions as Taylor series in powers of $1/R$.  From these 
series, we obtain in turn the pN-expansions of the $l=2, m=\pm2$ contributions 
to $\psi_0$, $\Psi$ and their first two radial derivatives and, finally, 
the pN-series of $H_{2,\pm2}^\textrm{ret}.$
}

\section{Numerical extraction and error analysis}

We describe in this section the way we numerically extract the pN-coefficients and check the accuracy with which they are determined.  We compute $\Delta U(R)$ for $R\,=\,1\times, \,3\times, \,5\times, \,8\times \, 10^{20}$ to $10^{29}$ in logarithmic intervals of $10$. From this data, after subtracting the known terms of Eq. (\ref{knownpN}), we match it to 
\begin{align} \label{fitseries}
   \sum_{n=5}\alpha_n\frac1{R^{n+1}} 
		+ \sum_{n=6} \beta_n\frac{\log R} {R^{n+1}} 
		+ \sum_{n=7} \gamma_n\frac{\log^2 R}{R^{n+1}} 
		+ \sum_{n=10}\delta_n\frac{\log^3 R}{R^{n+1}}
		+\cdots, .
\end{align}
The accuracy with which the coefficients are extracted depends on the number of terms in the series. The fit is done in $Mathematica$ and the number of terms in Eq. (\ref{fitseries}) chosen to maximize the accuracy of the extracted coefficients is calculated as follows. Since, the coefficient extracted depends on the number of terms (say $k$) in the fitting series, we give another index $k$ to some n-pN term, say the first unknown coefficient 5.5-pN term, $\alpha^{k}$ (we omit the pN-index for simplicity for now). We then look at the fractional difference of the $|\alpha_{k\pm1}/\alpha_{k}-1|$ vs $k$, and the $k_0$ at which the fractional difference is minimum, we choose that coefficient. For further details we refer the reader to Sec (V) of \cite{sf3} where a similar fitting is done. The fitting procedure is done twice here - first to extract the new analytical pN-coefficients (the terms in Eq. (\ref{newpN}) minus Eq. (\ref{knownpN})), and then we subtract them from data to do another fit to extract the coefficients in Table \ref{table}.

\section{Discussion}
\label{sect:disc}

In \cite{ltbw} it was established that a relation exists between, on the one hand, coefficients in the pN expansion of the red-shift variable and, on the other hand, coefficients in the expansion of the pN binding energy and angular momentum for the binary system; for explicit results, see, for example, Eq.~(2.50a-d) and (4.25a-d) in \cite{ltbw}. Subsequently, using essentially Eqs.~(2.40), (4.19) and (4.23) in \cite{ltbw}, Le Tiec at al.~(see \cite{LeTiec:2011dp}) transformed these relations to obtain, for arbitrary pN order, elegant expressions for the energy and angular momentum directly in terms of the self-force red shift variable and its first derivative; see, in particular, Eqs.~(4a-b) in \cite{LeTiec:2011dp}.

Self-force extensions of the pN binding energy and angular momentum have been long sought after, since they were known to have the potential to contribute to the effective-one-body (EOB) formulation (see \cite{Buonanno:1998gg, Damour:2000we}) of the binary inspiral problem --- mimicking, as far as possible, the reduced mass form of the Newtonian problem, but in a fully four dimensional, space-time setting.  Thus, in a follow-up paper to \cite{LeTiec:2011dp}, Barausse et al. \cite{Barausse:2011dq} found a very compact result, expressing the relevant EOB function directly in terms of the self force variable alone;  see their Eq.~(2.14) for this important relation, subsequently also reported in \cite{Akcay:2012ea}.

There is a very clear synergy between self-force results, and their applications in pN and EOB work, and knowledge of our new results will have an immediate impact though the application of the relations discussed in the previous two paragraphs.  Since the completion of our calculation, a corresponding computation has been performed to directly evaluate the 5.5pN coefficient through conventional pN analysis, in which it is known to arise from a tails-of-tails contribution.  
The ensuing result, as reported in a companion paper \cite{BFW5.5}, is in exact agreement with the 5.5pN term in our \Deqn{newpN}.

\noindent {\bf Note}:  The works cited in this section express $\Omega$ as $x^{3/2}\!/(M+\mathfrak{m})$ rather than our $R^{3/2}\!/{M}$, and use $z(x)=1/U(R)$ as the red shift variable.  The notation used throughout the rest of this paper was first introduced by Detweiler \cite{detweiler08}.

\noindent{\bf Note added:} At 1:15 pm (GMT) on December 9th, 2013 we received email notification from Nathan Kieran Johnson-McDaniel \cite{NKJM} that $\alpha_5$ could be represented as $$\frac{205680256 + 7342080\gamma - 31680075\pi^2 + 28968960\log(2) - 13996800\log(3)}{403200}.$$ An equivalent result, and an exact expression for $\alpha_6$, have subsequently appeared in \cite{BD2}. It has since been possible to show that our numerical results for $\beta_7$
and $\beta_8$ can be represented by $$\beta_7 = \frac{5163722519}{5457375} - \frac{109568}{525}\gamma - \frac{219136}{525}\log(2)$$ and $$\beta_8 = \frac{769841899153}{496621125} + \frac{108064}{2205}\gamma + \frac{5361312}{11025}\log(2) - \frac{852930}{2205}\log(3).$$  An explanation of these results and the methods used to obtain them will be discussed in a forthcoming paper \cite{JmSW}.


\begin{acknowledgments}
We are indebted to Alexandre Le Tiec for pointing out \cite{BiniDamour} for the value of 4-pN coefficient.  This work was supported by NSF Grant PHY 1001515 to UWM, PHY 0855503 to UF, European Research Council Starting Grant No. 202996 to WIS and the European Research Council under the European UnionÕs Seventh Framework Programme (FP7/2007-2013)/ERC grant agreement no. 304978 to UoS.  BFW acknowledges sabbatical support from the CNRS through the IAP, where part of this work was carried out.
\end{acknowledgments}

\bibliography{sf5}
\end{document}